\documentclass[superscriptaddress,amsmath,amssymb,aps,prb,twocolumn,nofootinbib,nolongbibliography]{revtex4-2}
\usepackage{graphicx}
\usepackage{graphics}
\usepackage{dcolumn}          
\usepackage{bm}               
\usepackage{upgreek}
\usepackage{booktabs} 				
\usepackage{tabularx}
\usepackage{array}
\usepackage[colorlinks=true,urlcolor=blue,breaklinks=true]{hyperref}

\usepackage{xcolor}

\usepackage{orcidlink}

\usepackage{xr}

\usepackage{siunitx}
\usepackage{stackengine}
\usepackage{svg}
\usepackage{graphicx,subfigure}
\usepackage{placeins}
\usepackage{bibunits}

\definecolor{green2}{RGB}{15, 117, 19}

\usepackage{titlesec}

\begin{document}

\title{Superconductive coupling and Josephson diode effect in selectively-grown topological insulator based three-terminal junctions}

\author{Gerrit Behner\,\orcidlink{0000-0002-7218-3841}}
\email{g.behner@fz-juelich.de}
\affiliation{Peter Gr\"unberg Institut (PGI-9), Forschungszentrum J\"ulich, 52425 J\"ulich, Germany}
\affiliation{JARA-Fundamentals of Future Information Technology, J\"ulich-Aachen Research Alliance, Forschungszentrum J\"ulich and RWTH Aachen University, Germany}

\author{Abdur Rehman Jalil
\,\orcidlink{0000-0003-1869-2466}}
\affiliation{Peter Gr\"unberg Institut (PGI-9), Forschungszentrum J\"ulich, 52425 J\"ulich, Germany}
\affiliation{JARA-Fundamentals of Future Information Technology, J\"ulich-Aachen Research Alliance, Forschungszentrum J\"ulich and RWTH Aachen University, Germany}

\author{Alina Rupp\,\orcidlink{0009-0009-6140-4387}}
\affiliation{Peter Gr\"unberg Institut (PGI-9), Forschungszentrum J\"ulich, 52425 J\"ulich, Germany}
\affiliation{JARA-Fundamentals of Future Information Technology, J\"ulich-Aachen Research Alliance, Forschungszentrum J\"ulich and RWTH Aachen University, Germany}

\author{Hans L\"uth\,\orcidlink{0000-0003-1617-3355}}
\affiliation{Peter Gr\"unberg Institut (PGI-9), Forschungszentrum J\"ulich, 52425 J\"ulich, Germany}
\affiliation{JARA-Fundamentals of Future Information Technology, J\"ulich-Aachen Research Alliance, Forschungszentrum J\"ulich and RWTH Aachen University, Germany}

\author{Detlev Gr\"utzmacher\,\orcidlink{0000-0001-6290-9672}}
\affiliation{Peter Gr\"unberg Institut (PGI-9), Forschungszentrum J\"ulich, 52425 J\"ulich, Germany}
\affiliation{JARA-Fundamentals of Future Information Technology, J\"ulich-Aachen Research Alliance, Forschungszentrum J\"ulich and RWTH Aachen University, Germany}

\author{Thomas Sch\"apers\,\orcidlink{0000-0001-7861-5003}}
\email{th.schaepers@fz-juelich.de}
\affiliation{Peter Gr\"unberg Institut (PGI-9), Forschungszentrum J\"ulich, 52425 J\"ulich, Germany}
\affiliation{JARA-Fundamentals of Future Information Technology, J\"ulich-Aachen Research Alliance, Forschungszentrum J\"ulich and RWTH Aachen University, Germany}

\hyphenation{}
\date{\today}

\begin{abstract}
\noindent The combination of an ordinary s-type superconductor with three-dimensional topological insulators creates a promising platform for fault-tolerant topological quantum computing circuits based on Majorana braiding. The backbone of the braiding mechanism are three-terminal Josephson junctions. It is crucial to understand the transport in these devices for further use in quantum computing applications. We present low-temperature measurements of topological insulator-based three-terminal Josephson junctions fabricated by a combination of selective-area growth of $\mathrm{Bi_{0.8}Sb_{1.2}Te_3}$ and shadow mask evaporation of Nb. This approach allows for the in-situ fabrication of Josephson junctions with an exceptional interface quality, important for the study of the proximity-effect. We map out the transport properties of the device as a function of bias currents and prove the coupling of the junctions by the observation of a multi-terminal geometry induced diode effect. We find good agreement of our findings with a resistively and capacitively shunted junction network model. 
\end{abstract}

\maketitle

\section{Introduction}
\begin{bibunit}[apsrev4-1]
Three-dimensional topological insulators (3D TIs) are a class of materials which recently raised a lot of interest due to its promising applicability in the field of topological quantum computing \cite{Hasan_2010,Ando_2013,Breunig_2021}. The material class exhibits strong spin-orbit coupling. This in turn leads to a band inversion in the bulk electronic band structure. As a consequence, gapless surface states appear, which are protected by time-reversal symmetry. Proximizing a topological insulator nanoribbon with an s-type superconductor and aligning a magnetic field along the nanoribbon gives rise to Majorana zero modes \cite{Lutchyn_2018}. Braiding of these Majorana zero modes is the essential computation operation in topological quantum computing \cite{Nayak_2008,Alicea_2012,Hyart_2013,Sarma_2015,Aasen_2016}. For this process multi-terminal structures are necessary in which the superconducting phase of the different electrodes needs to be adjusted.  The three-terminal Josephson junction therefore represents an important building block for these networks \cite{Cook_2011,Cook_2012,Manousakis_2017,de_Juan_2019,Legg_2021,Legg_2022,Heffels_2023}. It is crucial to understand the transport in these devices for further use in topological quantum computing applications. Generally, hybrid devices with multiple connections lead to rich physics in terms of transport properties, with a huge parameter space to be probed \cite{Pankratova_2020}. 

In recent years, the field of multi-terminal Josephson junctions and their unique properties have attracted a lot of interest in the scientific community,  e.g., the emergence of $n$-1 dimensional topological properties from an $n$-terminal Josephson junction made from conventional superconductor or the study of the synthetic Andreev band structure in the two-dimensional phase space \cite{Riwar_2016,Yokoyama_2015,Xie_2017,Xie_2018,Coraiola_2023}. While multi-terminal Josephson junctions have extensively been studied in epitaxially grown semiconductor-superconductor hybrid structures \cite{Coraiola_2023,Graziano_2020,Graziano_2022,Gupta_2023,Coraiola_2024}, not much has been reported on these devices in the field of topological materials. A flux-controlled three-terminal junction based on $\mathrm{Bi_2Te_3}$, revealed the opening and closing of a minigap \cite{Yang_2019, Pal_2022}. Furthermore, a three-terminal junction based on the topological insulator $\mathrm{Bi_4Te_3}$ was investigated, which did not show the expected signatures of the multi-terminal Josephson effect, but rather those of a resistively shunted network of two Josephson junctions \cite{Koelzer_2023}. 

Recently, the superconducting diode effect has attracted a lot of attention \cite{Nadeem_2023}. A characteristic of the diode effect is that the magnitude of the critical supercurrent depends on the direction in which the current is driven. The diode effect occurs when both inversion and time-reversal symmetry are broken. For Josephson junctions with a semiconducting \cite{Baumgartner_2022,Turini_2022,Costa_2023,Lotfizadeh_2024} or topological insulator \cite{Lu_2023} weak link, this can be accomplished by the presence of spin-orbit coupling in conjunction with an external magnetic field for the time-reversal symmetry breaking. Alternatively, the inversion symmetry can be broken by the device layout itself. This can be achieved, for example, by a superconducting quantum interference device (SQUID), where each of the two junctions of the interferometer has a different current phase relation \cite{Souto_2022}. More recently, the asymmetry in a multi-terminal Josephson junction has led to a diode effect, either by keeping one of the superconducting electrodes floating \cite{Gupta_2023} or by phase biasing using superconducting loops connecting pairs of electrodes in the junction \cite{Coraiola_2024}.

We present low-temperature measurements of three-terminal $\mathrm{Bi_{0.8}Sb_{1.2}Te_3}$ Josephson junctions fabricated by a combination of selective-area growth and shadow mask evaporation \cite{Sch_ffelgen_2019,Jalil_2023}. This approach allows for the in-situ fabication of Josephson junctions with very high interface transparency, important for the study of the superconducting proximity-effect.  The transport properties of the junction are mapped out as a function of bias current and magnetic field. The bias current maps show several interesting transport phenomena, e.g. an extended superconducting area and multiple Andreev reflections, indicating the successful fabrication of a fully-coupled three-terminal junction. The measured results on the junction appear to be in good agreement with a resistively and capacitively shunted junction model, but also reveal the influence of intrinsic asymmetries and their effect on transport in the junctions. The coupling of the junctions is emphasized by the observation of a multi-terminal geometry induced diode effect as a result of an externally applied magnetic field. 

\section{Experimental Methods}

\begin{figure}[h!]
    \centering
    \includegraphics[width=0.99\linewidth]{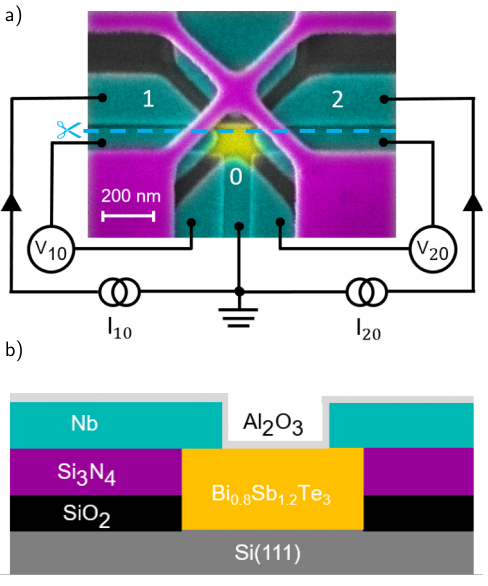}
    \caption{\textbf{a)}  Scanning electron micrograph of the device. The electronic setup used for applying current and measuring voltage is also shown. The terminal names (0,1,2) that are used for naming the single junctions are indicated in red. The blue dashed line indicates a lincut through the layerstack which is presented in b). \textbf{b)} Schematic illustration of the line cut (dashed blue line) in the layer stack making up the device.}
    \label{fig:Sample_Collage}
\end{figure}

The samples is fabricated using a combination of selective-area growth and shadow mask evaporation \cite{Sch_ffelgen_2019,Jalil_2023}. This gives the posibility to prepare samples with arbitrary geometry and exceptional interface transparency between the topological insulator and the parent superconductor. \SI{10}{nm} of SiO$_2$ and \SI{25}{nm} of Si$_3$N$_4$ are deposited by thermal oxidation and plasma-enhanced chemical vapor deposition (PECVD), respectively, on a Si(111) wafer. Trenches in the shape of a T and a width of 100\,nm are etched into the stack by means of a resist process using electron beam lithography and reactive-ion etching (RIE). A second stack of \SI{300}{nm} SiO$_2$ and \SI{100}{nm} Si$_3$N$_4$ is deposited using PECVD. This stack is subsequently be used to define the bridge for shadow evaporation. To do so, the second Si$_3$N$_4$ layer is patterned into the shape of the bridge using a negative resist process and RIE. By etching the sample with hydroflouric acid in a second etching step the bridge is under-etched creating a suspended shadow mask above the trench. The TI growth takes place under rotation of the sample around its normal axis. This ensures homogeneous growth under the shadow mask. The Nb contacts are deposited in-situ. For this purpose, \SI{50}{nm} of Nb is deposited from an angle without rotation of the sample. The shadow mask then patterns the Josephson junction itself, without the need for etching. Finally, the sample is capped using a \SI{5}{nm} layer of Al$_2$O$_3$ to prevent oxidation. The coarse shape of the electrodes are defined ex-situ using an SF$_6$ RIE process without damaging the junction area or the nanoribbon at any point. Figure~\ref{fig:Sample_Collage} a) depicts a scanning electron microscopy (SEM) image of the studied device, while Fig.~\ref{fig:Sample_Collage} b) shows a line cut in the layer stack along the light blue dashed line in a). 

The sample characteristics were measured in a dilution refrigerator with a base temperature of $T \approx \SI{10}{mK}$. Figure~\ref{fig:Sample_Collage} a) shows a typical measurements configuration, where two current sources supply currents $I_\mathrm{10}$ and $I_\mathrm{20}$ from the left and right terminal, i.e. terminals 1 and 2, respectively, to the bottom electrode, i.e. terminal 0, respectively. Voltages $V_\mathrm{10}$ and $V_\mathrm{20}$ are measured accordingly. The voltages are measured in a quasi four-point measurement scheme. The differential resistance of the sample is measured using a lock-in amplifier by addition of a \SI{10}{nA} AC current to the applied DC current. For the measurements of the diode effect an out-of-plane magnetic field is applied. 

\section{Experimental results}

\begin{figure}[h!]
    \centering
    \includegraphics[width=0.99\linewidth]{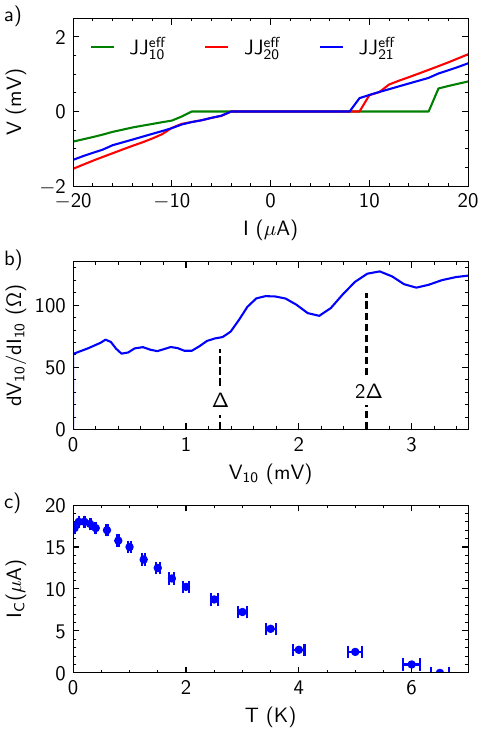}
    \caption{\textbf{a)} Current-voltage characteristics of the single junctions JJ$_{10}^\mathrm{eff}$, JJ$_{20}^\mathrm{eff}$, and JJ$_{21}^\mathrm{eff}$ in the device shown in Fig.~\ref{fig:Sample_Collage} a). \textbf{b)} Differential resistance of JJ$_{10}^\mathrm{eff}$ as a function of bias voltage $V_\mathrm{10}$. The position of the voltage bias for possible multiple Andreev reflections is indicated by dashed lines, where the first to are labeled $\mathrm{2\Delta}$ and $\Delta$. \textbf{c)} Temperature dependence of the critical current of JJ$_{10}^\mathrm{eff}$.}.
    \label{fig:DC_collage}
\end{figure}

\subsection{Two-terminal characteristics} \label{Sec:Two_Terminal}

First, the two-terminal transport properties of the multi-terminal device are characterized at zero magnetic field. This is done by recording the current-voltage ($IV$) curves between two terminals while the third is floating, effectively acting as a superconducting island (see Fig. \ref{fig:DC_collage} a)). If all measurements between pairs of superconducting electrodes show signatures of induced superconductivity, this implies that always a fraction of the bias current flows through the floating electrode. Note that the measurement results discussed below thus do not represent the properties of a single junction defined by the two contacted electrodes, but always include a contribution from the presence of the floating electrode. Nevertheless, it serves to gain a general picture of the junction properties. 

Here, we discuss the basic characteristics of the \textit{effective} Josephson junctions that are defined by contacting two of the three terminals, e.g. the bottom and the right superconducting arm. These terminals are indicated in red as 1 and 0 in Fig.~\ref{fig:Sample_Collage} a). Henceforward, the junction will be referred to as $\mathrm{JJ_{10}^\mathrm{eff}}$. Figure~\ref{fig:DC_collage} a) shows the DC-characteristics of all three effective junctions, i.e. $\mathrm{JJ_{10}^\mathrm{eff}}$, $\mathrm{JJ_{20}^\mathrm{eff}}$, and $\mathrm{JJ_{21}^\mathrm{eff}}$ at zero magnetic field. The junctions exhibit a hysteresis between the switching current $I_\mathrm{c}$ and re-trapping current $I_\mathrm{r}$. The current-voltage characteristics of a Josephson junctions can be described by a resistively and capacitively shunted junction (RCSJ) model. However, because of the coplanar junction geometry and the resulting very low junction capacitance we can rule out that the hysteresis is due to an overdamped junction characteristics. We rather attribute the hysteresis to Joule heating effects \cite{Courtois_2008}. Note that $\mathrm{JJ_{20}^\mathrm{eff}}$ and $\mathrm{JJ_{21}^\mathrm{eff}}$ exhibit similar switching currents slightly below $10\,\mu$A whereas $\mathrm{JJ_{10}^\mathrm{eff}}$ shows a critical current larger by almost a factor two. We attribute this difference in the switching currents to variations in the interface properties. The switching currents as well as the other characteristic parameters for all junctions are summarized in Table~\ref{tab:Junction-parameters}. In the following we will discuss the properties of the junction $\mathrm{JJ_{10}^\mathrm{eff}}$ as an example. 
\begin{table}[h]
    \centering
    \resizebox{0.5\linewidth}{!}{%
    \begin{tabular}{||c | c c c||}
         \hline
& JJ$_{10}^\mathrm{eff}$ & JJ$_{20}^\mathrm{eff}$ & JJ$_{21}^\mathrm{eff}$ \\ 
\hline\hline
$I_\mathrm{c}$ & 16 $\upmu$A & 9\,$\upmu$A & 8$\upmu$A \\ 
$R_\mathrm{N}$ & 113\,$\Omega$ & 138\,$\Omega$ & 131\,$\Omega$\\
$I_\mathrm{exc}$  & 19.1\,$\mu$A & 10.7\,$\mu$A & 12.6\,$\mu$A \\
$\tau$ & 0.94 & 0.87 & 0.89 \\
 \hline
    \end{tabular}}
    \caption{Single junction parameters: $I_c$ critical current, $R_\mathrm{N}$ normal state resistance, $\tau$ transparency.}
    \label{tab:Junction-parameters}
\end{table}

The excess current $I_\mathrm{exc}$ of the junctions is determined by a linear regression of the junctions ohmic behavior at bias voltages larger than $2\Delta$, with $\Delta$ the superconducting gap energy. Here, $2\Delta \approx 2.6\,$meV is determined from the critical temperature $T_\mathrm{c} \approx 8.5\,\mathrm{K}$ of the Nb film. The slope of the linear regression also determines the normal state resistance $R_\mathrm{N}$ of the junction, which is $\SI{113}{\Omega}$ for junction JJ$_{10}^\mathrm{eff}$. The excess current I$_\mathrm{exc} \approx \mathrm{19.1\,uA}$ is used to estimate the junction transparency $\tau$ of JJ$_{10}^\mathrm{eff}$. It can be gained by a fit to the Octavio-Tinkham-Blonder-Klapwijk (OTBK) model \cite{Octavio_1983,Flensberg_1988}. The junction transparency is a figure of merit to evaluate the interface quality. In fact, our junctions exhibits a large transparency, i.e. up to $\tau = 0.94$ for junction JJ$_{10}^\mathrm{eff}$. 

Figure~\ref{fig:DC_collage} b) shows the differential conductance as a function of bias voltage. Here, features in the differential conductance reveal subharmonic gap structures (SGS) \cite{Galletti_2017,Kunakova_2019,Jauregui_2018,Kim_2022,Ghatak_2018}. The signatures can in most cases be attributed to multiple Andreev reflections (MAR). They are expected to appear at voltages of $V = (2\Delta)/(en)$, where $e$ is the elementary charge and $n$ an integer \cite{Klapwijk1982,Octavio_1983}. The shape of the features are determined by the transport characteristics in the junction \cite{Averin_1995,Cuevas_2006,Schmitt_2022}. 
The positions of the voltage biases for possible multiple Andreev reflections are indicated by vertical dashed lines, where the first two are labeled $2\Delta$ and $\Delta$. The peak located at roughly \SI{1.7}{mV} bias voltage $V_\mathrm{10}$ could be a result of the 2$\Delta$ MAR resonance due to the induced gap in the TI weak link or in fact a result of a resistance change in on of the other junctions. The exact reason is hard to determine, as all junction are evidently coupled to each other, influencing the behaviour and measured results. Figure~\ref{fig:DC_collage} c) shows the temperature dependence of the switching current of JJ$_{10}^\mathrm{eff}$. The temperature dependence suggests behaviour of the weak link that is dominated by diffusive transport \cite{Sch_ffelgen_2019}.

\begin{figure*}[t]
    \centering
    \includegraphics[width=0.99\linewidth]{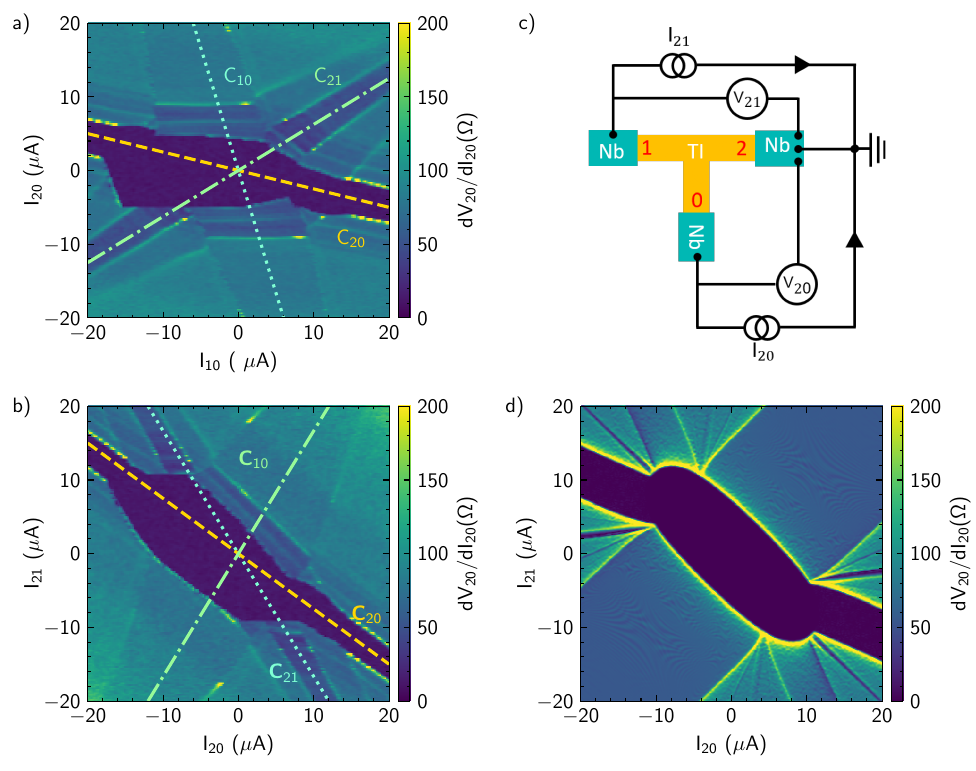}
    \caption{Differential resistance - current bias maps for different measurement configurations. \textbf{a)} Current bias map of the three-terminal junction measured in the setup shown in Fig.~\ref{fig:Sample_Collage} a). The bias map depicts the differential resistance of $dV_{20}/dI_{20}$ as a function of applied currents $I_{10}$ and  $I_{20}$. The three different lines (dotted, dashed, and dashdot) in the graph indicate the three regimes of compensated currents. The compensated currents are an effect mediated by the non-grounded Josephson junction coupling the two others. In this case, the coupling is mediated by $\mathrm{JJ_{12}}$. \textbf{b)} Current bias map of the measurement configuration shown in c). The junction determining the coupling between the two grounded junction is switched in order to increase the effects mediated by the coupling. The coupling junction in this case is $\mathrm{JJ_{20}}$. 
    \textbf{c)}  Schematic depiction of the RCSJ network model used for the simulated reproduction of the behaviour of the three-terminal Josephson junction.  \textbf{d)} Current bias map generated with the solution of the RCSJ network model. The simulation is carried out with values extracted from the measurements. A direct comparison of the experimental results in b) and the theoretically expected behavior in d) gives a reasonable qualitative agreement.  
    }
    \label{fig:BiasMaps}
\end{figure*}

\subsection{Three-terminal measurements}
\begin{figure}[]
    \centering
    \includegraphics[width=0.99\linewidth]{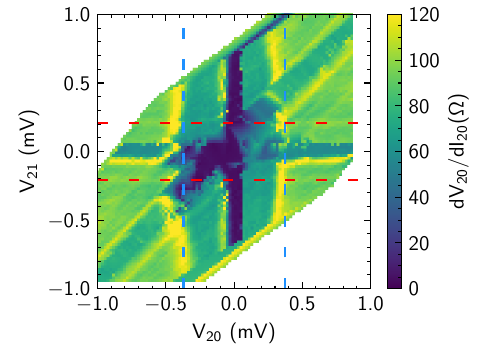}
    \caption{The bias map shown in Fig.~ \ref{fig:BiasMaps} c) but now with the differential resistance $dV_{20}/dI_{20}$ plotted as a function of the DC voltage drops $V_{20}$ and $V_{21}$. The horizontal and vertical lines at zero voltage can be attributed to the superconducting state between the corresponding pairs of terminals. The vertical and horizontal lines marked in blue and red we attribute to MAR in junctions JJ$_{20}$ and JJ$_{21}$, respectively.
    The diagonal line at $V_{20} = V_{21}$ corresponds to the superconducting state between the terminals 1 and 0.}
    \label{fig:Voltage}
\end{figure}

Due to the intrinsic asymmetry of the T-shape of our three-terminal junction, two different three-terminal configurations are probed. The first is schematically shown in Fig.~\ref{fig:Sample_Collage} a). Here, the coupling between the two driving junctions, i.e. from terminal 1 and 2 to terminal 0 at ground, respectively, is mediated by the junction connecting the left and right arm. The characteristics of the three-terminal junction are probed by measuring the differential resistance $dV_{20}/dI_{20}$ between terminals 2 and 0 for the respective configurations. The corresponding bias maps for $dV_{10}/dI_{10}$ are shown in the Supplemental Material.

Figure~\ref{fig:BiasMaps} a) shows the differential resistance $d V_{20}/dI_{20}$ measured in the configuration shown inFig.~\ref{fig:Sample_Collage} a) as a function of applied currents $I_{10}$ and $I_{20}$. The measurement outcome can be summarized by three prominent features \cite{Pankratova_2020}. The first and most distinct feature is an extended region of superconductivity in the center of the map indicated by the dark blue area. This corresponds to the zero-voltage state between terminal 2 and 0 of the three-terminal junction in the ($I_{10}$,\,$I_{20}$) plane. Since the switching and re-trapping current are different, the superconducting area is asymmetric with respect to the center of the current bias map. The second characteristic feature of a three-terminal junction is indicated by three lines in the bias map marked $C_{10}$, $C_{20}$, and $C_{21}$, respectively. They each represent a special combination of bias currents $I_{10}$ and $I_{20}$ for which either $V_{10}$, $V_{20}$, or $V_{21}$ is zero (see Fig \ref{fig:BiasMaps}.). This is a generic feature of three-terminal Josephson junction that exhibit dissipationless transport in all of its junctions. They are the result of a current being able to flow to ground by two different paths in the multi-terminal junction. For example, current $I_{20}$ can not only flow to ground directly via the junction JJ$_{20}$ formed between superconducting electrodes 2 and 0 but also take a detour through the other arms. Thus part of the current will also flow from terminal 2, via terminal 1 to ground.  For example, if $I_{20} \geq 0$ and $I_{10} \leq 0$, this will lead to a compensation of currents in junction JJ$_{20}$. As a result, the superconducting region in the bias map is extended along the diagonal. The slope of the extension is determined by the ratio of the normal state resistances of the junctions \cite{Pankratova_2020}. In Fig.~\ref{fig:BiasMaps} a) line $C_{20}$ therefore represents a compensation of currents in the probed junction, whereas line $C_{10}$ corresponds to a compensation of currents in junction $\mathrm{JJ_{10}}$ formed between electrode 1 and 0, visible as a reduced resistance. Line $C_{21}$ represents the compensation in the coupling junction $\mathrm{JJ_{21}}$. Here, the condition $\mathrm{sgn}(I_{10})=\mathrm{sgn}(I_{20})$ needs to be fulfilled so that the current components provided from terminal 2 and 1 flowing through $\mathrm{JJ_{12}}$ have opposite signs. 

In order to reveal the effect of the junction properties on the superconducting area a second configuration shown in Fig.~\ref{fig:BiasMaps} c) is investigated, where now $\mathrm{JJ_{20}}$ mediates the coupling between $\mathrm{JJ_{10}}$ and $\mathrm{JJ_{21}}$. Figure~\ref{fig:BiasMaps} b) shows the differential resistance $dV_{20}/dI_{20}$ as a function of the applied currents $I_{10}$ and $I_{21}$. Visible again is an extended superconducting region and the features due to compensating currents indicated by the lines $C_{10}$, $C_{20}$ and $C_{21}$. 

The third feature in the current bias maps are equipotential lines, which we attribute to MAR resonances in the junction according to the analysis of Pankratova \textit{et al.} \cite{Pankratova_2020}. Figure~\ref{fig:Voltage} shows the differential resistance $dV_{20}/dI_{20}$ as a function of the DC voltages $V_{20}$ and $V_{21}$ recorded in the measurement configuration shown in Fig.~\ref{fig:BiasMaps} c). The equipotential lines visible in the current bias color maps do indeed appear as lines of constant voltage for the respective junctions. We attribute them to MAR resonances in the sample. However, it is difficult to assign the respective lines to a fixed value of $V = (2\Delta)/(en)$, but it is reasonable to assume that the lines correspond to higher order MAR resonances.  
\begin{figure*}[ht]
    \centering
    \includegraphics[width=0.9\linewidth]{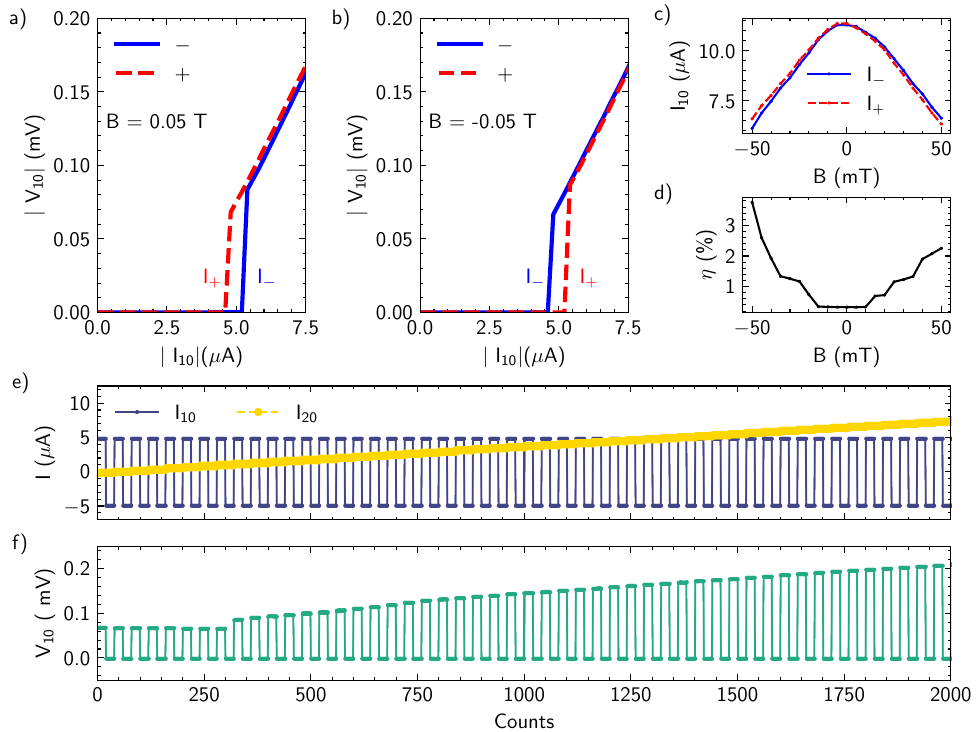}
    \caption{Diode effect in a three-terminal Josephson junction. \textbf{a)} Current-voltage characteristics of the effective junction JJ$^\mathrm{eff}_{10}$ for negative and positive bias currents at an applied out of plane magnetic field of \SI{50}{mT}. In this case, the critical current is larger for positive bias currents.  \textbf{b)} Corresponding set of IV characteristics after inversing the magnetic field to \SI{-50}{mT} with the critical current for negative bias currents being larger. \textbf{c)} Switching currents $I_+$ and $I_-$ as a function of magnetic field for positive and negative bias currents, respectively. \textbf{d)} Diode efficiency $\eta$ as a function of magnetic field. \textbf{e)} Periodic switching of current $I_{10}$ while linearly increasing the control terminal bias $I_{20}$. \textbf{f)} Response of JJ$^\mathrm{eff}_{10}$ to an alternating current $I_{10}$ and a linearly increasing current $I_{20}$ shown in e). A field of 50\,mT is applied.}
    \label{fig:Diode}
\end{figure*}
\subsubsection{Simulation}
Neglecting multiple couplings, our three-terminal Josephson junction can be described by a model that connects three Josephson junctions, i.e., JJ$_{10}$, JJ$_{20}$, and JJ$_{21}$, in a triangular network.  A schematics of the network is provided in the Supplemental Material. We employed the simulation approach presented in the work of Gupta \textit{et al.} \cite{Gupta_2023} to determine the relevant junction parameters. The $IV$ characteristics in the two-terminal measurements presented in Sect.~\ref{Sec:Two_Terminal} are used to extract first estimates of the critical currents and normal state resistances. In a second step, these values are adjusted in a semi-qualitative manner until a satisfactory agreement with the experimental data is obtained. The resulting bias map of the simulated three-terminal Josephson junction, corresponding to the measurements shown in Fig.~\ref{fig:BiasMaps} c), is shown in Fig.~\ref{fig:BiasMaps} d). An approximate matching of the critical current area as well as the slope  and position of the arms is achieved. The corresponding parameters, i.e. critical currents and normal state resistances for each individual junction are presented in the Supplemental Material. It turns out that a good matching is obtained by assuming a much higher critical current for junction JJ$_{10}$ compared to the other two junctions. In general, the critical currents of the individual junctions in the network are smaller than the values obtained from the two-terminal measurements because the bypass currents are not included. Since the junction capacitances are neglected and Joule heating is not implemented the simulation fails to reproduce the hysteresis of the device, which leads to a slight shift from zero in the experimental data. Since in the simulation effects are not included, the experimentally observed tapering of the three arms/slope for the higher current regimes cannot be reproduced. The tapering is a result of the dynamic reduction of the critical current due to Joule heating.
\subsubsection{Diode effect}
Due to the breaking of inversion symmetry by the device layout, three-terminal Josephson junctions are expected to exhibit a diode effect when exposed to a perpendicular magnetic field \cite{Gupta_2023}. Figure~\ref{fig:Diode} a) shows the current-voltage characteristics of JJ$_{10}^\mathrm{eff}$ with terminal 2 left floating. The current from terminal 1 to terminal 0 was increased from zero in either positive or negative direction to investigate the switching current for both current directions. By applying a magnetic field of \SI{50}{mT}, it is found that the switching current for positive bias currents ($I_{+}$) is smaller than that for negative currents ($I_{-}$). As can be seen in Fig.~\ref{fig:Diode} b), reversing the magnetic field to \SI{-50}{mT} causes the polarity of the diode effect to switch, with $I_{+}$ now being larger than $I_{-}$. 

In order to quantify the performance of the Josephson diode its efficiency is defined as follows: $\eta = \delta I_c / (I_+ + |I_-|)$, where $\delta I_c= (I_+ - |I_-|)$.
At \SI{\pm 50}{mT} we get a diode efficiency $\eta$ of about $0.04$ for JJ$_{10}^\mathrm{eff}$. For a fully symmetric three-terminal Josephson junction, the diode efficiency should follow a $\Phi_0/2$ periodicity, where $\Phi_0=h/(2e)$ is the magnetic flux quantum and $h$ is the Planck constant \cite{Gupta_2023}. A maximum diode efficiency $\eta_\mathrm{max}$ of 0.28 is expected at an applied external flux of $\Phi_0/4$. For our device, the magnitude of the magnetic field corresponding to $\Phi_0/4$ can be calculated from the junction size estimated from the SEM micrograph shown in Fig. \ref{fig:Sample_Collage} a). Assuming a junction area of $100 \times 100$\,nm$^2$, one obtains a magnetic field of $\approx\SI{50}{mT}$, corresponding to the field applied in our experiment. The experimentally determined value  at $\SI{\pm 50}{mT}$ of $\eta \approx 0.04$ is considerably smaller than the expected $\eta_\mathrm{max}$. We attribute the smaller value to the asymmetric layout of our multi-terminal Josephson junctions and to the unbalanced switching currents. Figure~\ref{fig:Diode} c) displays the critical currents $I_{-}$ and $I_{+}$ as a function of magnetic field with the evolution of the diode efficiency $\eta$ shown in Fig.~\ref{fig:Diode} d). It can be seen, that around \SI{0}{mT} both currents $I_{-}$ and $I_{+}$ are equal yielding $\eta=0\,\%$. When increasing the magnetic field into the respective direction, the difference between $I_{-}$ and $I_{+}$ becomes larger yielding increased efficiency.  

The diode characteristics of the single junction under application of an alternating current is depicted in the supplemental material. Here we focus on the characteristics determined by the three terminal layout. By biasing the second terminal with a current $I_{02}$ it is possible to control the diode characteristics \cite{Gupta_2023}. This is illustrated in Fig.~\ref{fig:Diode} e) and f), where the Josephson bias current $I_{10}$ is periodically switched between $\pm \SI{5}{\mu A}$ while the voltage drop $V_{10}$ is recorded. At the same time, the control bias current $I_{20}$ is ramped from 0 to \SI{15}{\mu A}. The voltage drop of JJ$_{10}^\mathrm{eff}$ recorded in Fig.~\ref{fig:Diode} f) shows that at zero control bias current there is already a diode effect, with a superconducting state at $I_{10}= \SI{-5}{\mu A}$ and a voltage drop of $\SI{0.075}{mV}$ at $\SI{+5}{\mu A}$. As long as the junction JJ$_{20}$ is in the superconducting state, i.e. up to $I_{20}=\SI{1}{\mu A}$, the diode characteristics do not change. Beyond this value, however, the voltage drop of the diode makes a sudden jump and increases approximately linearly. For negative diode current biases, the diode remains in the superconducting state up to the maximum control current of $\SI{15}{\mu A}$. Closer inspection reveals that at a current $I_{20}$ of about $\SI{2.5}{\mu A}$ the slope of the linear increase in $V_{10}$ decreases slightly. We attribute this to the fact that the two other junctions JJ$_{20}$ and JJ$_{21}$ are in the resistive state at the respective bias current.

\section{Conclusion}

In conclusion, we have fabricated a three-terminal Josephson junction in-situ, where the weak link material consists of a selectively grown topological insulator. The superconducting Nb electrodes are defined by shadow evaporation. This approach allows the fabrication of arbitrarily shaped junctions for future quantum computing applications, in particular structures for braiding Majorana zero modes. The single junctions that make up the device are of very high quality, with high transparency and evidence of multiple Andreev reflections. The three-terminal junction exhibits all the characteristics of a fully coupled multi-terminal Josephson junction. The observed slight asymmetries in the switching currents are attributed to the intrinsic asymmetry of the T-shape junction and to variations in the interface transparency. The bias current maps of the differential resistance show all features expected for multi-terminal Josephson junctions where all junctions interact in the superconducting state. The proof of principle now allows for more sophisticated experiments to detect signatures of topological superconductivity. By leaving one terminal floating and applying an external magnetic field, a diode effect is observed where the switching current depends on the direction of the bias current. As a next step, connecting superconducting electrodes by a superconducting loop to achieve phase bias would allow for an efficient control of the diode characteristics, offering a great potential for superconducting electronic circuits \cite{Coraiola_2024}.\\ 

\section{Acknowledgments}

We thank Herbert Kertz for technical assistance, and Florian Lentz and Stefan Trellenkamp for electron beam lithography. We thank Mohit Gupta for providing the code for the solution of the RCSJ network model. We are grateful for fruitful discussions with Kristof Moors, Roman Riwar and Peter Sch\"uffelgen. This work was partly funded by the Deutsche Forschungsgemeinschaft (DFG, German Research Foundation) under Germany’s Excellence Strategy - Cluster of Excellence Matter and Light for Quantum Computing (ML4Q) EXC 2004/1 – 390534769 as well as financially supported by the Bavarian Ministry of Economic Affairs, Regional Development and Energy within Bavaria’s High-Tech Agenda Project "Bausteine für das Quantencomputing auf Basis topologischer Materialien mit experimentellen und theoretischen Ansätzen" (grant no. 07 02/686 58/1/21 1/22 2/23).

\putbib[bu1.bbl]  
\end{bibunit}
\clearpage
\widetext

\titleformat{\section}[hang]{\bfseries}{\MakeUppercase{Supplemental Note} \thesection:\ }{0pt}{\MakeUppercase}
\begin{bibunit}[apsrev4-1]
\setcounter{section}{0}
\setcounter{equation}{0}
\setcounter{figure}{0}
\setcounter{table}{0}
\setcounter{page}{1}
\renewcommand{\thesection}{\arabic{section}}
\renewcommand{\thesubsection}{\Alph{subsection}}
\renewcommand{\theequation}{S\arabic{equation}}
\renewcommand{\thefigure}{\arabic{figure}}
\renewcommand{\figurename}{Supplemental Figure}
\renewcommand{\tablename}{Supplemental Table}
\renewcommand{\bibnumfmt}[1]{[S#1]}
\renewcommand{\citenumfont}[1]{S#1}

\begin{center}
\textbf{Supplemental Material: Superconductive coupling and Josephson diode effect in selectively-grown topological insulator based three-terminal junctions}
\end{center}

\section{Bias maps}
\begin{figure*}[]
    \centering
    \includegraphics[width=0.9\linewidth]{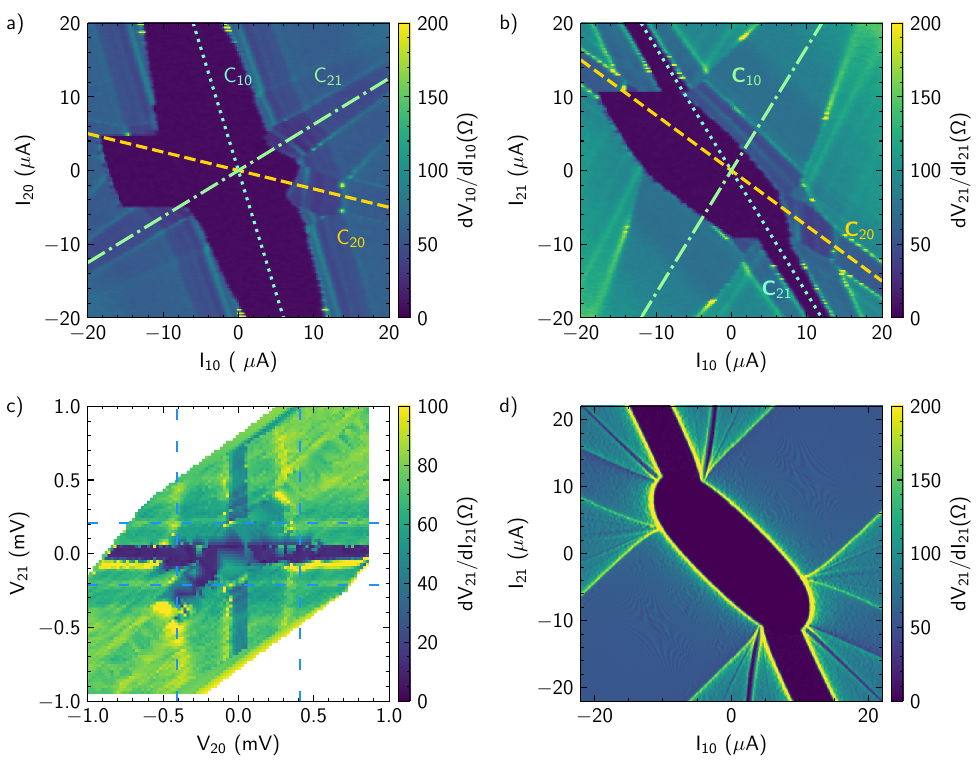}
    \caption{\textbf{a)} Differential resistance of $dV_{10}/dI_{10}$ as a function of applied currents $I_{10}$ and  $I_{20}$ measured in the setup shown in Fig.~1 a) in the main text. The three different lines $C_{10}$, $C_{20}$, and $C_{21}$ indicate the three regimes of compensated currents. \textbf{b)} Current bias map of $dV_{21}/dI_{12}$ for the measurement configuration show in Fig.~3 c) in the main text. \textbf{c)}  Differential resistance $dV_{21}/dI_{21}$ shown in b) now plotted as a function of the DC voltage drops $V_{20}$ and $V_{21}$. \textbf{d)} Current bias map generated with the solution of the RCSJ network model corresponding to the measurements shown in b).}
    \label{fig:BiasMaps_Supp}
\end{figure*}

Here, we present the corresponding counterparts to all current or voltage bias maps shown in the main text. Supplemental Figure~\ref{fig:BiasMaps_Supp} a) shows the differential resistance $dV_{10}/dI_{10}$ measured between terminals 1 and 0 as a function of applied currents $I_{10}$ and $I_{20}$. In contrast to Fig.~3 a) in the main text, the superconducting region extends along $C_{10}$ corresponding to the compensation of currents in junction JJ$_{10}$. The superconducting region of JJ$_{20}$ appears as a region of reduced resistance along $C_{20}$, while the compensation of currents in JJ$_{21}$ shows up as a region of reduced resistance along $C_{21}$. The same is true for Supplemental Figure~\ref{fig:BiasMaps_Supp} b), which shows the corresponding counterpart to the map in Fig.~3 b) given in the main text with the differential resistance $dV_{21}/dI_{21}$ plotted as a function of the applied currents $I_{10}$ and $I_{21}$. Here, the superconducting region extends along $C_{21}$ corresponding to the compensation of currents in the junction JJ$_{21}$. The compensating currents leading to supercurrents in junctions JJ$_{20}$ and JJ$_{10}$ appear as a region of reduced resistance along $C_{20}$ and $C_{10}$, respectively. Supplemental Figure~\ref{fig:BiasMaps_Supp} c) shows the differential resistance $dV_{21}/dI_{21}$ as a function of the DC voltages $V_{20}$ and $V_{21}$. Here, the region of zero resistance extends along the horizontal line corresponding to $V_{21}=0$. Similar to Fig.~4 in the main text, the lines of constant voltage due to multiple Andreev reflections are marked with horizontal and vertical dashed blue lines, respectively. Supplemental Figure~\ref{fig:BiasMaps_Supp} d) shows the corresponding counterpart to the simulation data for the experimental data given in Supplemental Figure~\ref{fig:BiasMaps_Supp} b).

\section{Simulation}
\begin{figure*}[h]
    \centering
    \includegraphics[width=0.5\linewidth]{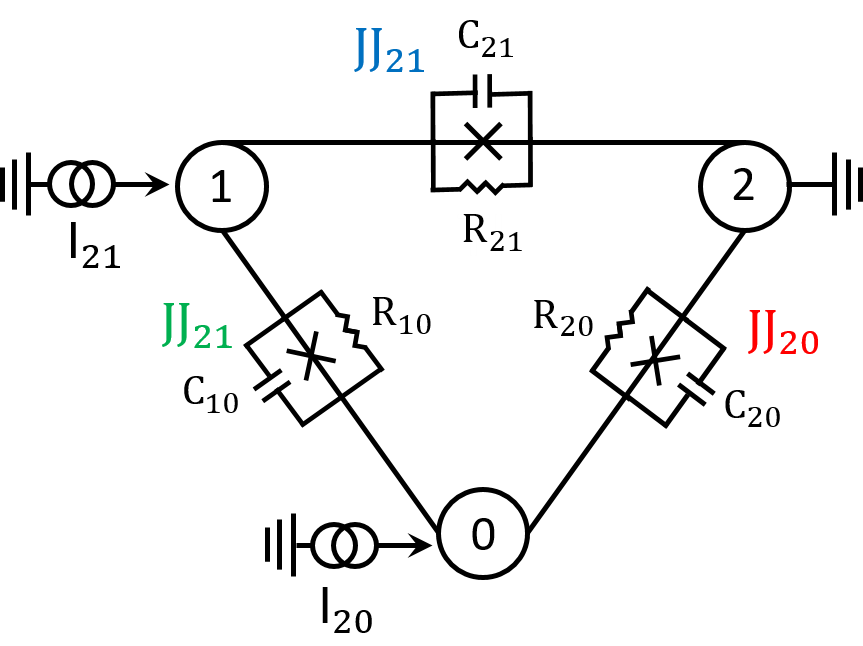}
    \caption{Effective transport model used for the simulations shown in the main section. The model is based on a network of three JJs described by a coupled resistively and capacitively shunted junction models The respective parameters are provided in Table \ref{tab:Simulation-parameters}.}
    \label{fig:SimParam}
\end{figure*}

This section provides details about the simulation and the corresponding parameters presented in the main section. Supplemental Figure~\ref{fig:SimParam} depicts the network of coupled resistively and capacitively shunted junctions. The three terminals are marked 0, 1, and 2. The corresponding junctions and the parameters are marked JJ$_{10}$, JJ$_{20}$, and JJ$_{21}$, respectively. A summary of the parameters used is given in Supplemental Table~\ref{tab:Simulation-parameters}. Both critical current I$_\mathrm{c}$ and the normal state resistances R$_\mathrm{N}$ are chosen smaller than the measured values. This is necessary to reproduce the features of the measurement, since the measured parameters are only the effective values and a consequence of the effective junction model, e.g. JJ$_{10}^\mathrm{eff}$, explained in the main section. The capacitance for each junction is set to \SI{5}{fF}. As shown by Graziano et al. \cite{Graziano_2020}, a small capacitance leads to an elliptical shape of the superconducting region, as observed in our experiments. For our junction layout, a small capacitance is expected because the shape of the junction represents a coplanar rather than a plate capacitor, as in the case of classical superconductor-insulator-superconductor (e.g., Al/AlO$_x$/Al) junctions. The capacitance of the junctions can be estimated by extracting the McCumber parameter $\beta_C$ as follows:
\begin{equation}
    \frac{I_\mathrm{c}}{I_\mathrm{r}} \approx \frac{4}{\pi}\cdot\frac{1}{\sqrt{\beta_C}}
\end{equation} \; , 
\begin{equation}
    \beta_C = \frac{2\pi}{\Phi_0}I_C R^2C.
\end{equation}
These considerations yield estimation for the capacitances of (C$_{10}^\mathrm{eff}$, C$_{20}^\mathrm{eff}$, C$_{21}^\mathrm{eff}$) = (\SI{3.9}{fF}, \SI{5.9}{fF}, \SI{5.8}{fF}), proving \SI{5}{fF} to be a reasonable choice for the capacitance used in the simulation. The same is true, when calculating the co-planar capacity of our device layout in a classical way. A more detailed look on the simulation approach used in this paper can be found in the supplementary material of reference \cite{Graziano_2022}. 

\begin{table}[h]
    \centering
    \resizebox{0.23\linewidth}{!}{%
    \begin{tabular}{||c | c c c||}
         \hline
& JJ$_{10}$ & JJ$_{20}$ & JJ$_{21}$ \\ 
\hline\hline
$I_\mathrm{c}$ & 8 $\upmu$A & 4\,$\upmu$A & 3.5$\upmu$A \\ 
$R_\mathrm{N}$ & 100\,$\Omega$ & 105\,$\Omega$ & 105\,$\Omega$\\
$C$  & 5\,fF & 5\,fF & 5\,fF \\
 \hline
    \end{tabular}}
    \caption{RCSJ simulation parameters: $I_\mathrm{c}$ critical current, $R_\mathrm{N}$ normal state resistance, $C$ capacitance.}
    \label{tab:Simulation-parameters}
\end{table}

\section{Magnetic field dependence}

\begin{figure*}[h]
    \centering
    \includegraphics[width=0.9\linewidth]{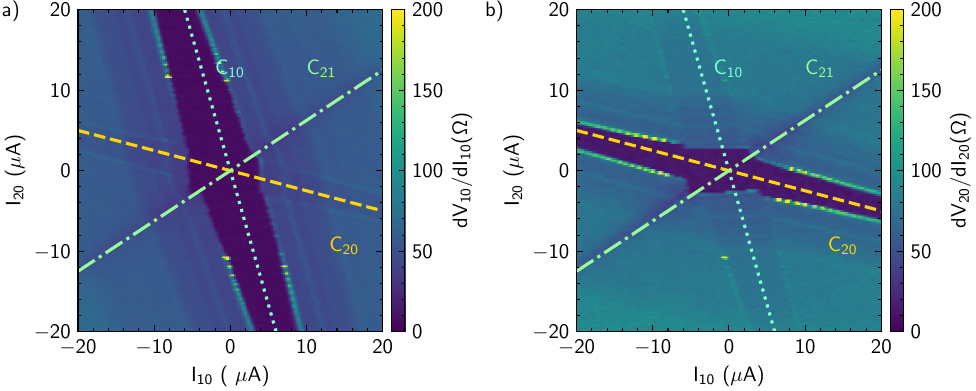}
    \caption{\textbf{a)} Differential resistance $dV_{10}/dI_{10}$ as a function of currents $I_{10}$ and $I_{20}$ under application of an perpendicular magnetic field of 50\,mT. The diagonal lines $C_{10}$, $C_{20}$, and $C_{21}$ indicate the features due to compensation currents in the junctions. \textbf{b)} Corresponding current bias map with $dV_{20}/dI_{20}$ plotted instead of $dV_{10}/dI_{10}$.}
    \label{fig:BiasMaps_OOP_Supp}
\end{figure*}

In this section we present the effect of an perpendicular magnetic field on the transport in the junctions. Supplemental Figures~\ref{fig:BiasMaps_OOP_Supp} a) and b) show the bias map corresponding to Supplemental Figure~\ref{fig:BiasMaps_Supp} a) and Fig.~3 a) in the main text, respectively, when a magnetic field of \SI{50}{mT} is applied. As expected, the junction properties decrease with the application of the field. The features that characterize a multi-terminal junction are strongly suppressed. One finds that the central regions of the Josephson supercurrent are shrunk, with the individual critical current reduced by more than a factor of two. The diagonal features resulting from compensating currents in the junctions, characterized by the lines $C_{10}$, $C_{20}$ and $C_{21}$, are almost vanished from the bias map. The lines attributed to the presence of multiple Andreev reflections are barely visible.  

\begin{figure*}[h]
    \centering
    \includegraphics[width=0.9\linewidth]{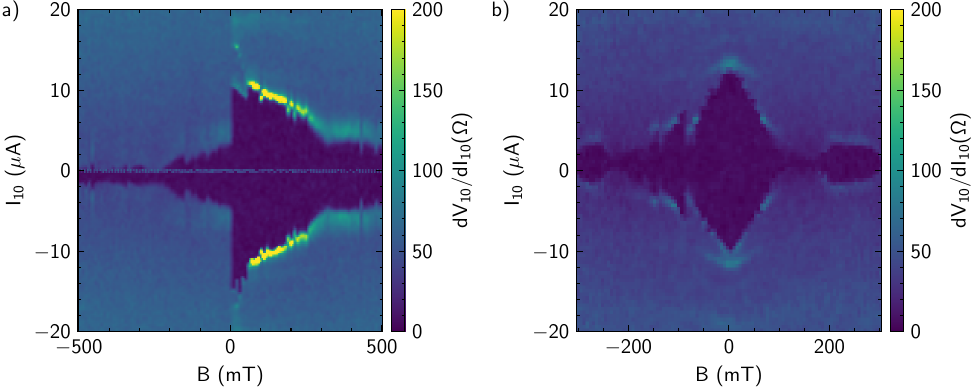}
    \caption{Differential resistance $dV_{10}/dI_{10}$ of JJ$_{10}$ as a function of an perpendicular magnetic field $B$ and current $dI_{10}$, recorded in two different ways. \textbf{a)} The magnetic field is swaped from zero in the respective direction while measuring the current voltage characteristics at every individual magnetic field point. \textbf{b)} The differential resistance $dV_{10}/dI_{10}$ is recorded as a function of bias current for a magnetic field starting from \SI{-300}{mT} and then consecutively increased to \SI{300}{mT}.}
    \label{fig:Fraunhofer_Supp}
\end{figure*}

Supplemental Figure~\ref{fig:Fraunhofer_Supp} shows the differential resistance $dV_{10}/dI_{10}$ as a function of an perpendicular magnetic field and bias current $I_{10}$. 
The map shown in Fig.~\ref{fig:Fraunhofer_Supp} a) is taken starting from \SI{0}{mT} and sweeping up to the maximum field of -500/500\,mT, while recording individual $IV$ curves at the respective field points. The dark blue regions represent the superconducting state. While the map shows signs of periodic behavior for positive fields, corresponding to a Fraunhofer interference pattern, the negative side shows a completely different asymmetric behavior. Judging from the size of the device extracted from the scanning electron microscopy image shown in Fig.~1 a) in the main text, the expected magnetic field periodicity should be in the order of $250-300\,$mT. The local minimum of the critical current found at about \SI{300}{mT} is in the expected range. We attribute the strongly suppressed supercurrent on the negative magnetic field side to a trapped vortex. With Nb as a type II superconductor, the device is susceptible to flux trapping when an perpendicular magnetic field is applied. This trapped flux exposes the junction to an effective magnetic field, resulting in an abrupt change in junction behavior.

In the Supplemental Figure~\ref{fig:Fraunhofer_Supp} b) $dV_{10}/dI_{10}$ is plotted starting at \SI{-300}{mT} and then successively sweeping up to \SI{300}{mT}, recording individual $IV$ curves at the respective field points. The behavior of the junction appears to be completely different, with a central lobe of superconductivity extending to about $-100/100\,$mT, but drifting to an asymmetric behavior for higher fields. We attribute this to two effects, i.e. a non-uniform current distribution at the junctions and the flux trapping that occurs at large fields.  

\section{Diode effect}

\begin{figure*}[h]
    \centering
    \includegraphics[width=0.79\linewidth]{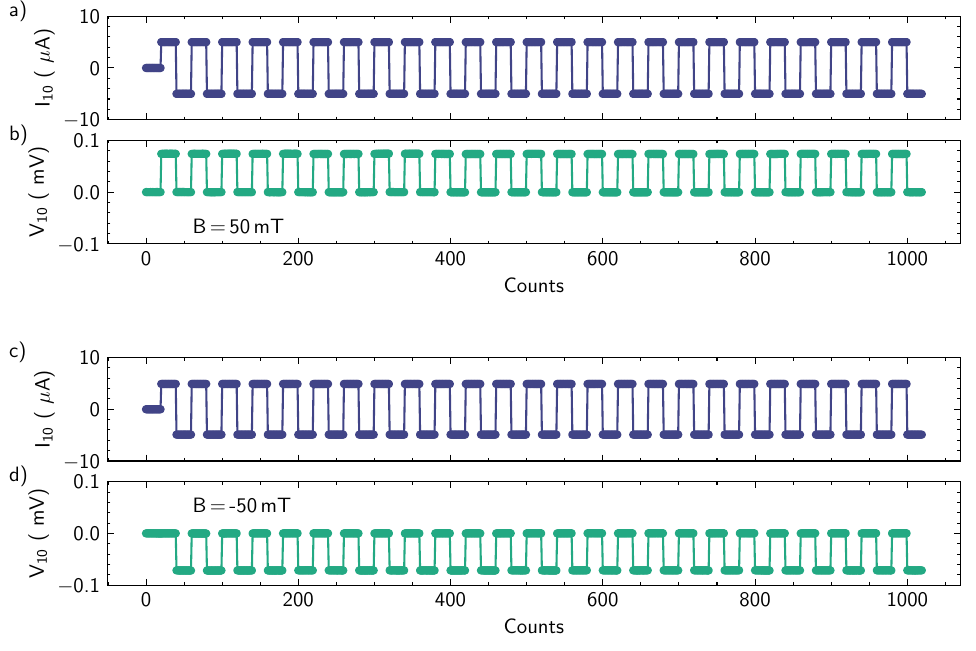}
    \caption{Demonstration of rectification of a current signal with amplitude $\mathcal{A}=\SI{5}{\mu A}$ in an perpendicular magnetic field. \textbf{a)} Applied square wave bias current signal with amplitude $\mathcal{A}=\SI{5}{\mu A}$. \textbf{b)} Corresponding voltage drop $V_{10}$ at the junction JJ$^\mathrm{eff}_{10}$ for a magnetic field of \SI{50}{mT}. \textbf{c)} \textbf{d)} Square wave bias current with amplitude $A=\SI{5}{\mu A}$ and corresponding voltage drop $V_{10}$ for a magnetic field of \SI{-50}{mT}.}
    \label{fig:Diode_JJ01}
\end{figure*}

Supplemental Figure~\ref{fig:Diode_JJ01} b) shows the response of the junction JJ$^\mathrm{eff}_{10}$ to an applied square-wave pulse at an applied magnetic field of \SI{50}{mT}. It demonstrates the rectification of a current signal with amplitude $\mathcal{A}=\SI{5}{\mu A}$. The amplitude $\mathcal{A}$ of the square wave is chosen so that $|I_-|<\mathcal{A}<I_+$. For negative currents the device remains superconducting, i.e. $V=\SI{0}{mV}$, while for positive currents the device becomes resistive with a voltage drop of $V_{10} \approx\SI{70}{\mu V}$. No punch-through errors are observed over tens of cycles. This demonstrates the stability of the effect even though the diode efficiency is rather low.

Supplemental Figure~\ref{fig:Diode_JJ01} c) and d) are the corresponding counterparts of a) and b) at an applied field of $-50\,$mT. As described in the main section on the diode effect, the polarity of the effect should be reversed when the magnetic field is reversed. This is demonstrated here by inverting the rectification of a current signal with amplitude $\mathcal{A}=\SI{5}{\mu A}$. In contrast to the measurements shown in Supplemental Figure~\ref{fig:Diode_JJ01} b), the device remains superconducting for positive currents, i.e. $V=\SI{0}{mV}$, while it becomes resistive for negative currents with a voltage drop of $V \approx\SI{-70}{\mu V}$. Similarly, no punch-through errors are observed over the entire measurement cycle.

\begin{figure*}[h]
    \centering
    \includegraphics[width=0.79\linewidth]{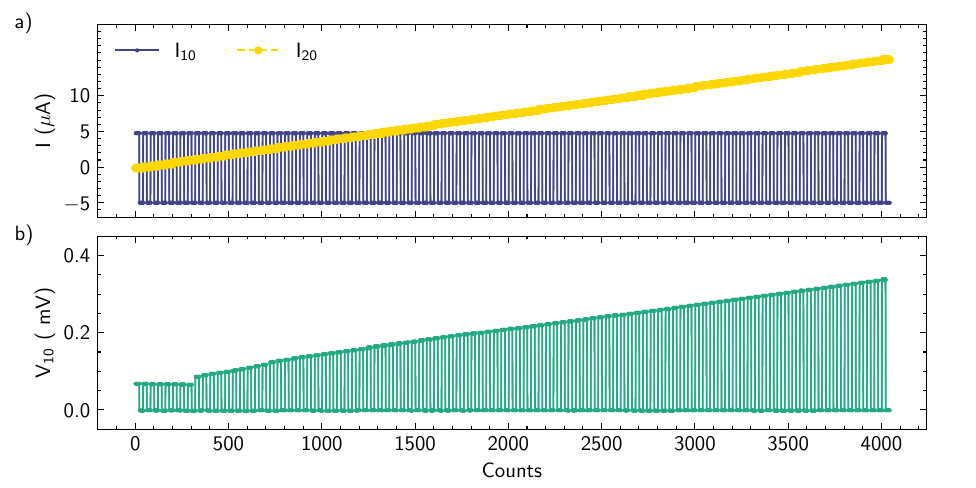}
    \caption{Rectification of a current signal in JJ$^\mathrm{eff}_{10}$ while biasing the second terminal with a current $I_{20}$ at an applied field of 50\,mT. \textbf{a)} Square wave signal with an amplitude $\mathcal{A}=\SI{5}{\mu A}$ in blue and the linearly increasing second terminal bias current $I_{20}$ in yellow. \textbf{b)} Voltage response $V_{01}$ of junction JJ$^\mathrm{eff}_{10}$ to the applied signal shown in a).}
    \label{fig:Diode_JJ01_NonLin_Supp}
\end{figure*}

Supplemental Figure~\ref{fig:Diode_JJ01_NonLin_Supp} a) and b) show the complete measurement of the rectification of a current signal of amplitude $\mathcal{A}=\SI{5}{\mu A}$ applied from terminal 1 to 0 while the second terminal is biased with a current $I_{20}$. The measurement is taken at a magnetic field of \SI{50}{mT}. The second terminal bias current $I_{20}$ is ramped up to a total current of \SI{15}{\mu A} while $I_{10}$ is periodically switched between $\pm \SI{5}{\mu A}$. For negative currents the device remains superconducting, i.e. $V=\SI{0}{mV}$, while for positive currents the device becomes resistive. The slight bends at a second terminal bias of about $~\SI{1}{\mu A}$ and \SI{2.5}{\mu A} correspond to the junctions JJ$_{20}$ and JJ$_{21}$, which become resistive. For $I_{20}$ larger than about \SI{3}{\mu A}, a linear increase of the voltage $V_{10}$ with increasing $I_{20}$ is observed. The key features of the measurement are discussed in the main text. No punch-through errors are recorded throughout the measurement.
\section{Differential IV-Curve}
\begin{figure*}[h]
    \centering
    \includegraphics[width=0.79\linewidth]{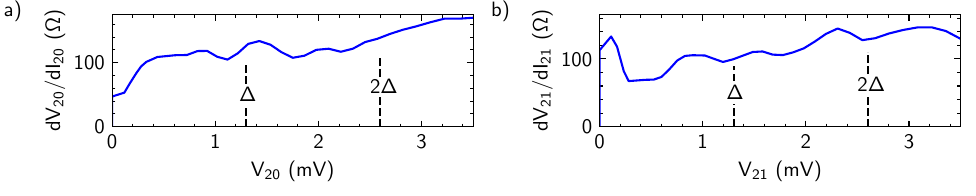}
    \caption{\textbf{a)} and \textbf{b)} Differential resistances of the junctions JJ$^\mathrm{eff}_{20}$ and JJ$^\mathrm{eff}_{21}$ as a function of the DC voltages $V_{20}$ and $V_{21}$. The positions of possible MAR are marked with $2\Delta$ and $\Delta$, respectively.}
    \label{fig:DiffIV_Supp}
\end{figure*}

Supplemental Figure~\ref{fig:DiffIV_Supp} a) and b) shows the differential resistance of JJ$^\mathrm{eff}_{20}$ and JJ$^\mathrm{eff}_{21}$ as a function of the DC voltages $V_{20}$ and $V_{21}$, respectively. Similar to Fig.~1 b) in the main text, the positions of possible multiple Andreev reflections are marked with vertical dashed lines labeled with $2\Delta$ and $\Delta$, respectively.

\putbib[bu2.bbl]  
\end{bibunit}
\end{document}